\def\rfr#1{eq. (\ref{#1})}

\def\derp#1#2{\rp{\partial{#1}}{\partial{#2}}}
\def\dert#1#2{\frac{{{d}}{#1}}{{{d}}{#2}}}
\def\Om{\mathit{\Omega}}
\def\virg#1{``#1"}

\def\bb#1#2#3{\bibitem[\protect\citeauthoryear{#1}{#2}]{#3}}
\def\eqi{\begin{equation}}
\def\eqf{\end{equation}}
\def\eqia{\begin{eqnarray}}
\def\eqfa{\end{eqnarray}}
\def\rp#1#2{{#1\over#2}} \def\lb#1{\label{#1}}
\def\bb#1#2#3{\bibitem[\protect\citeauthoryear{#1}{#2}]{#3}}
\def\bds#1{\boldsymbol{#1}}
\def\kap{\bds{\hat{\xi}}}
\def\kx{\hat{\xi}_x}
\def\ky{\hat{\xi}_y}
\def\kz{\hat{\xi}_z}
\def\si{\sin I}
\def\ci{\cos I}
\def\so{\sin\omega}
\def\co{\cos\omega}
\def\sO{\sin\Om}
\def\cO{\cos\Om}
\def\su{\sin u}
\def\cu{\cos u}

\documentclass{aastex}
\usepackage{url}
\usepackage{amsmath,amsthm,amscd,amssymb}
\usepackage{latexsym,wasysym}
\usepackage{graphicx,epsfig}

\RequirePackage{color}

\newcommand{\emaila}{lorenzo.iorio@libero.it}

\begin{document}

\title{Observational constraints on spatial anisotropy of $G$ from orbital motions}
%\slugcomment{Receive}
%% Running heads
%\shorttitle{On the imprint}
\shortauthors{L. Iorio}

\author{Lorenzo Iorio\altaffilmark{1} }
\affil{Ministero dell'Istruzione, dell'Universit\`{a} e della Ricerca (M.I.U.R.)-Istruzione\\
Fellow of the Royal Astronomical Society (F.R.A.S.)\\
International Institute for Theoretical Physics and
High Mathematics Einstein-Galilei\\
 Permanent address for correspondence: Viale Unit\`{a} di Italia 68, 70125, Bari (BA), Italy}

\email{\emaila}

\begin{abstract}
A phenomenological anisotropic variation $\Delta G/G$ of the Newtonian gravitational coupling parameter $G$, if real, would affect the orbital dynamics of a two-body gravitationally bound system in a specific way. We analytically work out the long-term effects that such a putative modification of the usual Newtonian inverse-square law would induce on the trajectory of a test particle orbiting a central mass. Without making any a-priori simplifying assumptions concerning the orbital configuration of the test particle, it turns out that its osculating semi-major axis $a$,  eccentricity $e$,  pericenter $\varpi$ and  mean anomaly $\mathcal{M}$  undergo long-term temporal variations, while the inclination $I$ and the node $\Om$ are left unaffected. Moreover, the radial and the transverse components of the position and  the velocity vectors $\bds r$ and $\bds v$ of the test particle experience non-vanishing changes per orbit, contrary to the out-of-plane ones. Then, we compute our theoretical predictions for some of the major bodies of the solar system  by orienting the gradient of $G(\bds r)$ towards the Galactic Center and keeping it fixed over the characteristic timescales involved. By comparing our calculation
to the latest observational determinations for the same bodies, we infer  $\Delta G/G\leq 10^{-17}$ over about 1 au. Finally, we  consider also the Supermassive Black Hole hosted by the Galactic Center in Sgr A$^{\ast}$ and the main sequence star S2 orbiting it in about 16 yr, obtaining just $\Delta G/G\leq 10^{-2}$ over 1 kau.
\end{abstract}

%\keywords{gravitation-celestial mechanics}
%\keywords{black hole physics-Galaxy:center-relativity-techniques: radial velocities}
\keywords{Experimental studies of gravity; Experimental tests of gravitational theories; Modified theories of gravity; Orbital and rotational dynamics}
PACS: 04.80.-y; 04.80.Cc; 04.50.Kd;  96.25.De
\section{Introduction}
The possibility that the Newtonian coupling parameter $G$ may experience macroscopic spacetime variations ranging from laboratory to cosmological scales has been investigated both theoretically \citep{Scia53,Brans61,Dick62,Long74,Ger88,Linde90,deSab92,Mel94a,Mel94b,Capoz96,Mel96,Capoz97,Capoz98,Drink98,Fisch98,Bar01,Dani01,Kra01,Mur01,Uzan03,deSab04,Capoz05,Clif05,Ham05,Bail06,Bro06,Garcia07,Berto08,Capoz08,Adel09,Uzan09,Ahma010,Kost011,Uzan011} and experimentally/observationally  \citep{Wago70,Vinti72,Ulrich74,Long76,Warb76,Mikkel77,Anderson78,Blinni78,Hut81,Chan82,Kislik83,Chan84,Gill87,Burg88,Tal88,Kra92,Iz93,Paik94,Berto96,Gill97,Gatz02,Ger02,Ger04,Unni02a,Unni02b,Adel03,Lo03,Abra04,Bar05,Bou05,Kono05,Garcia07,Ior07,Berto09,New09,Li09,Lamm11,Piedi011,Uzan011}  since the early insights by \citet{Milne35,Milne37}, \citet{Dir37} and \citet{Jordan37,Jordan39}.

In this paper we  deal with possible smooth anisotropic spatial variations of $G$, i.e. we consider the case $G=G(\bds r)$ \textcolor{black}{from a purely phenomenological point of view. We stress that in our analysis we do not rely upon any specific theoretical scheme encompassing such a spatial variability of $G$: the interested reader may want to consult the previously cited specialized literature.}
Quite generally,  we  express a putative anisotropic dependence of $G$ on the spatial coordinates  by parameterizing it with a gradient $\bds\nabla G$ along a fixed direction in space $\kap$ as
\eqi G(\bds r)\simeq G_0+\bds\nabla G_0\bds\cdot\bds r,\lb{gvar}\eqf
with
\eqi \bds\nabla G_0 = \left|\bds\nabla G_0\right|\kap.\lb{gvar2}\eqf The subscript $\virg{0}$ in \rfr{gvar} and \rfr{gvar2} refers to quantities evaluated at the origin of the spatial coordinates which, in our case, coincides with a generic body of mass $M$ acting as source of the gravitational field. By means of \rfr{gvar} we are assuming that the putative  variations of $G$ are rather smooth over the spatial extensions considered.
A change of $G$ like that of \rfr{gvar} is usually absent in the standard alternative metric theories of gravity treated within the Parameterized Post-Newtonian (PPN) framework \citep{Will93} in which $G$ may depend on the velocity $\bds V$ of the frame in which the experiments are performed with respect to a preferred frame \citep{Will71}; see, e.g., \citet{Vinti72}, \citet{Dam94} and \citet{Will93}for  analyses of the orbital consequences of such kind of anisotropies. A spatial anisotropy of $G$ depending on the angle between the line of interaction of two gravitating bodies and a reference direction with respect to distant stars  was experimentally investigated by\footnote{They claimed to have measured an anisotropy as large as $5.4\times 10^{-4}$ with a torsion balance.} \citet{Ger02,Ger04} in a series of Earth-based laboratory investigations which were subsequently critically analyzed by \citet{Unni02a}. \citet{Unni02a} remarked that, in general,  spatial anisotropies of $G$ may occur depending on how nearby masses and their distribution can affect the gravitational interaction between two bodies, and also because of preferred frame effects.
If the modification of the gravitational interaction depends on the gravitational potential generated by other masses \citep{Brans61,Scia53} in a somewhat Machian fashion, then the most distant ones dominate and the spatial anisotropy is expected to be small \citep{Unni02a}. It is, then, reasonable to expect that the Galaxy yields the most important contribution to the anisotropy by assuming $\kap$ directed towards the Galactic  Center (GC) \citep{Unni02a}.

In the framework of our parameterisation of \rfr{gvar} and \rfr{gvar2}, such a  scenario may offer, in principle, interesting
observational perspectives if suitable astronomical bodies are chosen. Indeed,
if we insert \rfr{gvar} in the usual expression of the Newtonian inverse-square law,
a small modification of it occurs
\eqi\bds A=-\rp{M\left(\bds\nabla G_0\bds\cdot\bds r\right)\bds r}{r^3}=-\rp{M\left|\bds\nabla G_0\right|\left(\kap\bds\cdot\bds r\right)\bds r}{r^3}\lb{accel}.\eqf Now,
the ecliptic coordinates of the GC are \citep{Reid04}
\eqi
\begin{array}{lll}
\lambda_{\rm GC} & = & 183.15\ {\rm deg}, \\ \\
\beta_{\rm GC} & = & -5.61\ {\rm deg},
\end{array}
\eqf
so that the angle between $\kap$ and $\bds r$ for, say, a typical planet of the solar system is rather small. Thus,
\eqi \rp{\Delta G}{G}\sim \rp{\left|\bds\nabla G_0\right|r}{G_0}.\eqf \citet{Unni02a} argued that the order of magnitude of the Galactic-induced anisotropy is
\eqi\rp{\Delta G}{G}\sim \delta=\rp{GM_{\rm Gal}}{c^2 d}\sim 10^{-6},\eqf where $c$ is the speed of light in vacuum, $M_{\rm Gal}\sim 10^{12} M_{\odot}$ \citep{Batta05} is the mass of the Galaxy, and $d=8.28$ kpc \citep{Gille09} is the distance from the GC. In our picture, it would  naively be equivalent to  perturbing accelerations
\eqi A\lesssim 10^{-6}A_{\rm N},\eqf where the standard inverse-square Newtonian accelerations $A_{\rm N}$ are
\eqi 4\times 10^{-2}\ {\rm m\ s}^{-2}\leq A_{\rm N}\leq 4\times 10^{-6}\ {\rm m\ s}^{-2}\eqf for the major bodies of the solar system. In principle, perturbing accelerations as large as \eqi A\sim 10^{-8}-10^{-12}\ {\rm m\ s}^{-2}\eqf may have interesting observational consequences. We will analytically work out them in detail. Indeed,  relying upon simple order-of magnitude evaluations may be misleading since important factors of the order of $\mathcal{O}(e^j),\ j=1,2,\ldots$ or $\mathcal{O}(e^{-j}),\ j=1,2,\ldots$ in the usually small eccentricities $e$ of the bodies adopted as probes may be neglected. \citet{Unni02a} performed a preliminary calculation concerning the Earth-Moon system. They started from a certain value $\left|\dot G/G\right|\leq 4\times 10^{-12}$ yr$^{-1}$ \citep{Dick94} of the upper bound  in the fractional time change of $G$ obtained with the Lunar Laser ranging (LLR) technique. Then, \citet{Unni02a} stated that the same analysis can be useful as far as the spatial anisotropy of $G$ is concerned as well. They noticed that an anisotropic spatial variation of $G$ like that of \rfr{gvar} should exhibit a harmonic signal with the same approximate monthly periodicity of the orbital lunar motion as the line joining the Earth and the Moon sweeps out different directions with respect to the GC direction. Finally, since the accuracy with which it is possible to measure a periodic signal may be of the same order of, or better
than that for a secular trend,  \citet{Unni02a} concluded by inferring\footnote{It is so also because 1 year contains almost 12 lunar cycles.}
\eqi\rp{\Delta G}{G}\leq 4\times 10^{-12}.\lb{bound}\eqf Repeating the same reasonings with the latest results from LLR \citep{Will04,Mull07,Will09,Hof010} would yield
\eqi\rp{\Delta G}{G}\leq (1-0.4)\times 10^{-12},\eqf in neat disagreement the results by \citet{Ger02}. \citet{Ger02} pointed out that such a disagreement exists if it is assumed  that the $G$ anisotropy depends neither on the magnitude of the interacting masses nor on  the distance between them; \citet{Ger02} remarked that the masses and the distances involved in the analysis by \citet{Unni02a} drastically differ from those used by \citet{Ger02} in their experiment.

We propose to obtain much more accurate bounds than that in \rfr{bound} by calculating in detail all the orbital effects of \rfr{accel} which, actually, depends on the mutual distance between $M$ and the test body: we will assume that it is independent of their masses.  As far as other performed and/or proposed astronomical tests of $G(\bds r)$  are concerned \citep{Wago70,Vinti72,Ulrich74,Warb76,Mikkel77,Anderson78,Blinni78,Hut81,Kislik83,Burg88,Tal88,Abra04,Ior07,Berto09,Li09}, quite different explicit theoretical isotropic models and empirical approaches have been followed so far; in particular, an exponential Yukawa-type model and the third Kepler law have often been  adopted.

The paper is organized as follows. In Section \ref{calcoli} we analytically work out various orbital effects caused by a phenomenological dipole-type spatial variation of $G$ like that of \rfr{gvar} averaged over one orbital revolution of a test particle. In Section \ref{osservazioni} a comparison with the latest solar system planetary observations is made; we also consider the stellar system around the Galactic black hole. Section \ref{conclusioni} summarizes our findings and contain the conclusions.
\section{Calculation of the orbital effects}\lb{calcoli}
The orbital motions of test particles are, in principle, affected by \rfr{accel} whose effects can be worked out with standard perturbative techniques \citep{Brouw61,BeFa}.
By defining\footnote{Notice that $[\psi_0]={\rm L}^{-1}$, while $[G_0 M]= {\rm L}^3\ {\rm T}^{-2}$, as usual.}
\eqi \psi_0\doteq \rp{\left|\bds\nabla G_0\right|}{G_0},\ \mu_0\doteq G_0 M, \eqf
the radial component $A_R$ of  \rfr{accel}, evaluated onto the unperturbed Keplerian ellipse, is
\eqi A_R=-\rp{\psi_0 \mu_0\left(1+e\cos f\right)\left\{ \cu\left(\kx\cO+\ky\sO\right)+\su\left[\kz\si+\ci\left(\ky\cO-\kx\sO\right)\right] \right\}}{a\left(1-e^2\right)};\lb{acceR}\eqf
the transverse and out-of-plane components $A_T,A_N$ vanish. In \rfr{acceR} $a,I,\Om,u\doteq\omega+f,\omega,f$ are the osculating semi-major axis,  the inclination of the orbital plane to the reference $\{x,y\}$ plane, the longitude of the ascending node\footnote{It is an angle in the reference $\{xy\}$ plane counted from the reference $x$ direction to the line of the nodes, which is the intersection of the orbital plane with the reference $\{xy\}$ plane.},  the argument of latitude, the argument of pericentre\footnote{It is an angle in the orbital plane reckoning the point of closest approach with respect to the line of the nodes.} and the true anomaly\footnote{It is a time-dependent angle in the orbital plane determining the instantaneous position of the test particle with respect to the pericentre.}, respectively, of the test particle. Notice that \rfr{acceR} clearly shows that the putative anisotropic $G$ effect depends, among other things, on the distance between the two bodies  as well.
By assuming $\kap$ constant over one orbital revolution of the test particle, a straightforward first-order application of the Gauss perturbative equations \citep{Brouw61,BeFa} yields for  $a$,  $e$, the longitude\footnote{It is a \virg{dogleg} angle.} of the pericenter $\varpi\doteq\Om+\omega$, and  $\mathcal{M}$ the following non-vanishing long-term temporal rates of change
\eqi
\begin{array}{lll}
\dert a t &=& -\rp{2\psi_0 \mu_0}{aen}\left\{-\so\left(\kx\cO+\ky\sO\right) +\co\left[\kz\si +\ci\left(\ky\cO-\kx\sO\right)\right]\right\}, \\ \\
\dert e t & = & -\rp{\psi_0 \mu_0\left(1-e^2\right)}{a^2e^2n}\left\{-\so\left(\kx\cO+\ky\sO\right)+\co\left[\kz\si +\ci\left(\ky\cO-\kx\sO\right)\right]\right\}, \\ \\
\dert\varpi t & = & -\rp{\psi_0 \mu_0\left(1-e^2\right)}{a^2e^3n}\left\{\co \left(\kx\cO+\ky\sO\right) +
\so\left[\kz\si+\ci\left(\ky\cO-\kx\sO\right)\right]\right\}, \\ \\
\dert{\mathcal{M}} t & = & \rp{\psi_0 \mu_0\left(1-e^2\right)^{3/2}}{a^2e^3n}\left\{\co \left(\kx\cO+\ky\sO\right) +
\so\left[\kz\si+\ci\left(\ky\cO-\kx\sO\right)\right]\right\}.
\end{array}\lb{rates}
\eqf
%See \citet{Li09} for analogous calculations.
The inclination $I$ and the node $\Om$ are left unaffected.
Notice that the quantity $n$ entering \rfr{rates} is the unperturbed Keplerian mean motion, i.e. $n\doteq\sqrt{\mu_0/a^3}$.
The long-term rates of change of \rfr{rates} are exact in the sense that no a-priori assumptions concerning $\kap$ and the orbital configuration  of the test particle were made. All the formulas of \rfr{rates} are singular for $e\rightarrow 0$; however, it is just an unphysical artifact which can be cured by adopting the well-known non-singular elements \citep{Brouw61,Brou72}
\eqi
\begin{array}{lll}
h &\doteq & e\sin\varpi, \\ \\
k &\doteq & e\cos\varpi, \\ \\
l &\doteq & \varpi+\mathcal{M}.
\end{array}
\eqf
It is also important to remark that the validity of \rfr{rates} is not restricted to any specific reference frame, being, instead, quite general.

It is useful to work out the radial, transverse, and out-of-plane shifts over one orbital revolution  of the position and velocity vectors $\bds r$ and $\bds v$ as well.
They can be analytically worked out according to \citet{Caso93}. For the shifts $\Delta R,\Delta T,\Delta N$ of $\bds r$ we have
\eqi
\begin{array}{lll}
\Delta R & = & \rp{2\pi\psi_0 a^2\left(1-e^2\right)^2}{e^2}\left\{  -\so \left(\kx\cO+\ky\sO\right) +\co\left[ \kz\si +\ci\left(\ky\cO-\kx\sO\right)\right]\right\}, \\ \\
\Delta T &=& \rp{4\pi\psi_0 a^2\left(1-e^2\right)}{e^2}\left\{ \co\left(\kx\cO+\ky\sO\right)
+\so\left[\kz\si+\ci\left(\ky\cO-\kx\sO\right)\right]\right\}, \\ \\
\Delta N &=& 0.
\end{array}\lb{RTN}
\eqf
The shifts $\Delta v_R, \Delta v_T, \Delta v_N$ of $\bds v$ are
\eqi
\begin{array}{lll}
\Delta v_R & = &-\rp{2\pi\psi_0 a^2 n\left(1+e\right)^{3/2}}{e^2\sqrt{1-e}}\left\{
\co\left(\kx\cO+\ky\sO\right) + \so\left[\kz\si +\ci\left(\ky\cO-\kx\sO\right)\right]
  \right\}, \\ \\
\Delta v_ T & = & -\rp{2\pi\psi_0 a^2 n\sqrt{1-e^2}}{e^2}\left\{
-\so\left(\kx\cO+\ky\sO\right) +\co\left[\kz\si +\ci\left(\ky\cO-\kx\sO\right)\right]
\right\}, \\ \\
\Delta v_N & = & 0.
\end{array}\lb{vRvTvN}
\eqf
Concerning the singularities occurring for $e\rightarrow 0$, also for \rfr{RTN} and \rfr{vRvTvN} the same considerations as for \rfr{rates} hold.
\section{Confrontation with the observations}\lb{osservazioni}
Here we compare  our theoretical results of Section \ref{calcoli} with the latest observationally determined quantities pertaining the orbital motions of some of the major bodies of the solar system obtained by processing long data records of various types with different ephemerides \citep{Pit07,FiengaJournees010,Fie010,Pit010} in order to preliminarily infer upper bounds on $\psi_0$. In principle, one should explicitly include \rfr{gvar} in the dynamical force models fit to the observations and re-process the entire data set with such an $ad-hoc$ modified theory by varying the parameters to be estimated, the data, etc. Otherwise, the putative signal may be partly or totally absorbed in the estimation of, say, the initial state vectors. However, this is beyond the scopes of our work.

In computing the anomalous effects for different bodies, we  refer their orbital configurations to a heliocentric frame with mean ecliptic and equinox at the epoch J$2000$. In it
the unit vector pointing to the GC is
\eqi
\begin{array}{lll}
\kx & = & -0.993, \\ \\
\ky & = & -0.054, \\ \\
\kz & = & -0.097.
\end{array}\lb{versore}
\eqf
In Table \ref{errori} we quote reasonable evaluations for the secular variations of the semi-major axes $a$ and the eccentricities $a$ of the inner planets of the solar system for which the most accurate data are currently available. They were computed by dividing  the formal, statistical $1-\sigma$ errors in $a,e$ of the EPM2006 ephemerides \citet{Pit07} by the time interval $\Delta t=93$ yr covered by the observations used for constructing them.
\begin{table*}[ht!]
\caption{Formal uncertainties in the secular variations of the semi-major axes $a$ and the eccentricities $e$ of the inner planets of the solar system. They were obtained by dividing the formal errors in $a$ and $e$ in Table 3 of \protect{\citet{Pit07}} by the time interval $\Delta t=93$ yr (1913-2006) of the data records used for the EPM2006 ephemerides \protect{\citep{Pit07}}. The errors in $e$ were computed as $\sigma_e =\sqrt{\left(\derp e h\right)^2\sigma_h^2 + \left(\derp e k\right)^2\sigma_k^2}$. The realistic uncertainties may be up to one order of magnitude larger.
}\label{errori}
\centering
\bigskip
\begin{tabular}{lll}
\hline\noalign{\smallskip}
Planet & $\sigma_{\dot a}$ (m yr$^{-1}$)    & $\sigma_{\dot e}$ (yr$^{-1}$) \\
\noalign{\smallskip}\hline\noalign{\smallskip}
Mercury & $3\times 10^{-3}$  & $4.43\times 10^{-12}$ \\
Venus & $2\times 10^{-3}$  & $1.8\times 10^{-13}$  \\
Earth & $1\times 10^{-3}$  & $5\times 10^{-14}$ \\
Mars & $3\times 10^{-3}$  & $5\times 10^{-14}$ \\
\noalign{\smallskip}\hline\noalign{\smallskip}
\end{tabular}
\end{table*}
The realistic uncertainties may be up to one order of magnitude larger.
Table \ref{pericentri} displays the latest determinations of the corrections $\Delta\dot\varpi$ to the standard Newtonian/Einsteinian secular precessions of the perihelia of the planets of the solar system recently obtained with different ephemerides \citep{FiengaJournees010,Fie010,Pit010}.
\begin{table*}[ht!]
\caption{Estimated corrections $\Delta\dot\varpi$, in milliarcseconds per century (mas cty$^{-1}$), to the standard Newtonian-Einsteinian secular precessions of the longitudes of the perihelia $\varpi$ of the eight planets of the solar system plus Pluto determined with the EPM2008 \protect{\citep{Pit010}}, the INPOP08 \protect{\citep{Fie010}},  and the INPOP10a \protect{\citep{FiengaJournees010}} ephemerides. Only the usual Newtonian-Einsteinian dynamics was modelled, so that, in principle, the corrections $\Delta\dot\varpi$ \textcolor{black}{may} account for any other unmodelled/mismodelled dynamical effect. Concerning the values quoted in the third column from the left, they correspond to the smallest uncertainties reported by \protect{\citet{Fie010}}. Note the small uncertainty in the correction to the precession of the terrestrial perihelion, obtained by processing Jupiter VLBI data \protect{\citep{Fie010}}.
}\label{pericentri}
\centering
\bigskip
\begin{tabular}{llll}
\hline\noalign{\smallskip}
Planet & $\Delta\dot\varpi$  \protect{\citep{Pit010}}  & $\Delta\dot\varpi$  \protect{\citep{Fie010}} &  $\Delta\dot\varpi$  \protect{\citep{FiengaJournees010}} \\
\noalign{\smallskip}\hline\noalign{\smallskip}
Mercury & $ -4 \pm 5 $  & $ -10\pm 30$ & $ 0.2\pm 3$ \\
Venus & $ 24\pm 33$  & $-4\pm 6 $ & $ - $ \\
Earth & $ 6\pm 7$  & $ 0 \pm 0.016 $ & $ - $\\
Mars & $ -7\pm 7$  & $0\pm 0.2 $ & $ - $\\
Jupiter & $ 67\pm 93$  & $142\pm 156$ & $ - $\\
Saturn & $ -10\pm 15$ & $-10\pm 8$ & $ 0\pm 2$ \\
Uranus & $ -3890\pm 3900$  & $0\pm 20000$ & $ - $\\
Neptune & $ -4440\pm 5400 $  & $0\pm 20000$ & $ - $\\
Pluto & $ 2840 \pm 4510 $  & $-$ & $ - $\\
\noalign{\smallskip}\hline\noalign{\smallskip}
\end{tabular}
\end{table*}
Since  $\psi_0$ enters \rfr{rates} as a free, adjustable parameter, we select the values for it which make the putative anomalous effects of \rfr{rates} not larger than the empirically obtained bounds in Table \ref{errori} and Table \ref{pericentri}. As a result, we are able to obtain reasonable guesses concerning $\Delta G/G$ at different heliocentric distances by posing $\Delta G\sim\sigma_{\psi_0} \left\langle r\right\rangle$. The results are shown in Table \ref{DGG}.
\begin{table*}[ht!]
\caption{Upper bounds on the anisotropic percent variation $\Delta G/G$ inferred from  $\dot a,\dot e,\dot\varpi$  for the inner planets of the solar system. We posed $\Delta G\lesssim \sigma_{\psi_0} \left\langle r\right\rangle=\sigma_{\psi_0} a\left(1+e^2/2\right)$ for each planet, where $\sigma_{\psi_0}$ was obtained by comparing the theoretical predictions of \rfr{rates} to the uncertainties listed in Table \ref{errori} and Table \ref{pericentri}.
}\label{DGG}
\centering
\bigskip
\begin{tabular}{llll}
\hline\noalign{\smallskip}
Planet & $\left.\rp{\Delta G}{G}\right|_{\dot a}$    & $\left.\rp{\Delta G}{G}\right|_{\dot e}$ & $\left.\rp{\Delta G}{G}\right|_{\dot\varpi}$ \\
\noalign{\smallskip}\hline\noalign{\smallskip}
Mercury & $2\times 10^{-16}$  & $8\times 10^{-15}$ & $3\times 10^{-13}$\\
Venus & $9\times 10^{-18}$  & $1\times 10^{-18}$  & $1\times 10^{-17}$ \\
Earth & $1\times 10^{-17}$  & $2\times 10^{-18}$  & $3\times 10^{-18}$ \\
Mars & $4\times 10^{-16}$  & $3\times 10^{-16}$  & $3\times 10^{-15}$ \\
\noalign{\smallskip}\hline\noalign{\smallskip}
\end{tabular}
\end{table*}
Even by re-scaling the bounds obtained from Table \ref{errori} by one order of magnitude, the anisotropy of $G$ in the planetary regions of the solar system is very tightly constrained, being of the order of $10^{-15}-10^{-17}$.

%Similar constraints can be obtained from the Earth-Moon system whose range is nowadays known at a cm level or less after about 40 yr of operations %\citep{Will09,Hof010}. Indeed, by using the expression for $\Delta R$ in \rfr{RTN} for the radial perturbation it is possible to infer
%\eqi \left.\rp{\Delta G}{G}\right|_{\Delta R}\leq 4\times 10^{-17}\lb{limite}\eqf
%over $\Delta t=40$ yr, corresponding to about 480 lunar revolutions, with
%\eqi \psi_0^{-1}\geq 269\ {\rm Mpc}.\eqf
%The bound of \rfr{limite} is several orders of magnitude more stringent than the one by \citet{Unni02a}.

It may be interesting to consider a completely different astronomical scenario, both from the point of view of its components and of the distance scales  involved.
In Table \ref{tavolabuco} we quote the relevant physical and orbital parameters of the  system constituted by the Supermassive Black Hole (SBH) hosted by the GC in Sgr A$^{\ast}$ and  the main sequence star S2 orbiting it in about 16 yr at a distance of approximately 1 kau  \citep{Gille09}. In this case, the angular elements refer to a coordinate system whose reference $z$ axis is directed along the line of sight: the reference $\{x,y\}$ plane coincides with the plane of the sky\footnote{It is a plane tangent to the Celestial Sphere at the point where the object of interest is located.}, with the $x$ axis pointing towards the Celestial North Pole.
\begin{table*}[ht!]
\caption{Relevant physical and orbital parameters of the SBH-S2 system in Sgr A$^{\ast}$. The Keplerian orbital elements of S2 were retrieved from Table 1 of  \protect{\citep{Gille09}}. The figure for the gravitational parameter $\mu_0$ comes from a multi-star fit yielding $\mu_0=4.30\times 10^6\mu_{\odot}$ \protect{\citep{Gille09}}\textcolor{black}{: it yields a Schwarzschild radius as large as $r_g=0.084$ au}. The quoted value in m for the semi-major axis of S2  was obtained by multiplying  its angular value  $a=0.1246$ arcsec \protect{\citep{Gille09}} by the distance of the SBH $d=8.28$ kpc \protect{\citep{Gille09}}: it corresponds to $1031.69$ au, so that the orbital period of S2 is $P_{\rm b}=15.98\ {\rm yr}$. The uncertainty $\sigma_{\dot\omega}$ in the secular precession of S2 can naively be obtained by dividing the error in $\omega$, quoted in Table 1 of  \protect{\citep{Gille09}}, by the time interval $\Delta T$ covered by the observations used which is almost equal to $P_{\rm b}$.
}\label{tavolabuco}
\centering
\bigskip
\begin{tabular}{lllllll}
\hline\noalign{\smallskip}
$\mu_0$ (m$^3$ s$^{-2}$) & $a$ (m)  & $e$  & $I$ (deg) & $\Om$ (deg) & $\omega$ (deg) & $\sigma_{\dot\omega}$ (arcsec yr$^{-1}$)\\
\noalign{\smallskip}\hline\noalign{\smallskip}
$5.70\times 10^{26}$ & $1.54\times 10^{14}$ & $0.8831$ & $134.87$ & $226.53$ & $64.98$ & $182$\\
\noalign{\smallskip}\hline\noalign{\smallskip}
\end{tabular}
\end{table*}
The results of \rfr{rates} are applicable to S2 as well since its mass  is about five orders of magnitude smaller than that of the SBH: clearly, in this case it is $\kap=-\bds{\hat{z}}$.
\textcolor{black}{Concerning the use of \rfr{rates} in the Sgr A$^{\ast}$ scenario, one may wonder why an essentially Newtonian scheme is adopted instead of a general relativistic one. In principle, one could assume as
reference path a fully post-Newtonian one\footnote{\textcolor{black}{In doing so, it would be implicitly assumed that the effects due to a putative $G$ anisotropy are smaller than the post-Newtonian ones as well.}} \citep{Calura1,Calura2}, and work out the effects of a given
small extra-acceleration like \rfr{accel} with respect to it according to the perturbative
scheme set up by \citet{Calura1,Calura2}, which is a general relativistic generalization of another standard
perturbative approach based on the planetary Lagrange
equations \citep{BeFa}. Actually, it is, in practice, useless since the only addition with respect
to the orbital effects like, e.g., the precession of the pericenter, resulting from the
standard scenario would consist of further, small mixed GTR-perturbation orbital
effects, completely irrelevant in strengthening the bounds inferred. Viewed from a different point of view,  the ratio of the average distance $\overline{r}_{\rm S2}$ of S2 from the SBH to its Schwarzschild radius $r_g$ is, after all, as large as $1.7\times 10^4$.}

By using the expression for the putative precession of the stellar pericenter in \rfr{rates}, it can be compared to the present-day uncertainty in observationally determining its secular rate in Table \ref{tavolabuco}. In this case, the constraints on the $G$ anisotropy are very weak. Indeed, we have just
\eqi \psi_0^{-1}\geq 0.45\ {\rm pc},\eqf corresponding to
\eqi\rp{\Delta G}{G}\leq 1.5\times 10^{-2}\lb{micamale}\eqf over about 1 kau.
Note that the bound of \rfr{micamale} is tighter by one order of magnitude than that could be inferred by simply posing
\eqi\rp{\Delta G}{G}\lesssim \rp{\sigma_{\mu}}{\mu_0}=1.2\times 10^{-1},\eqf from $\sigma_{\mu}=0.50\times 10^{6}\mu_{\odot}$ \citep{Gille09}.

\section{Summary and conclusions}\lb{conclusioni}
We looked at  phenomenological anisotropic spatial variations $\Delta G/G$ of the Newtonian gravitational coupling parameter $G$, in the form $\Delta G=\bds\nabla G\bds\cdot\bds r$, and analytically worked out the impact that they may have on the trajectory of a test particle orbiting a central body of mass $M$. More specifically, we focussed on the cumulative orbital changes obtained perturbatively by averaging over one period of revolution of the test particle the effects due to $\Delta G/G$ on its path. As a result, the osculating semi-major axis $a$, the eccentricity $e$, the pericenter $\varpi$ and the mean anomaly $\mathcal{M}$ of the orbiter experience non-vanishing long-term changes which depend on the overall orbital geometry of the test particle and on the direction $\kap$ of the putative gradient $\bds \nabla G$. We analytically worked out the long-term variations per orbit $\Delta R,\Delta T,\Delta N$ and $\Delta v_R,\Delta v_T,\Delta v_N$ of the position and velocity vectors $\bds r$ and $\bds v$ of the test particle as well. We found that both the radial and the transverse components of $\bds r$ and $\bds v$ are affected by long-term changes per orbit, while the out-of-plane ones are left unaffected. We kept $\kap$ fixed during the integrations: moreover, no a-priori simplifying assumptions on $e$ and $I$ were assumed, so that our results are exact in this respect.

Then, we compared our theoretical predictions to  the most recently determined observational quantities for some of the major bodies of the solar system. By assuming that the dominant contribution to the hypothetical anisotropy of $G$ is due to the Galaxy, we took $\kap$ directed towards the Galactic Center, which has a small inclination with respect to the ecliptic. By using the heliocentric orbits of the inner planets
%and the orbital motion of the Moon about the Earth
we were able to constrain $\Delta G/G$ to a level of about $10^{-17}$ over $\sim 1$ au,  several orders of magnitude better than in previous analyses based on Lunar Laser Ranging only. We looked also at the star S2 orbiting the Galactic black hole at a distance of about 1 kau along a highly elliptical ellipse, but, in this case, we got just $\Delta G/G \lesssim 10^{-2}$.

\end{document}